\documentclass[twocolumn,twoside,slac_two]{revtex4}
\usepackage{graphicx}
\usepackage{fancyhdr}
\begin{document}

\title{The LAT Low-Energy technique for {\it Fermi} Gamma-Ray Bursts
  spectral analysis.}

\author{V. Pelassa, F. Piron}
\affiliation{LPTA, CNRS/IN2P3 - Universit\'e Montpellier 2}
\author{R. Preece, S. Guiriec}
\affiliation{NSSTC, MSFC - UAH}
\author{N. Omodei}
\affiliation{INFN, Pisa University}
\author{on behalf of the {\it Fermi} LAT and GBM collaborations}

\begin{abstract}
  {\it Fermi} Large Area Telescope (LAT) data analyses based on 
  event reconstruction and classification are so far restricted to
  events of measured energy larger than 100 MeV. We present a
  new technique to recover the signal from Gamma-Ray Bursts' (GRB) prompt
  emission between $\sim$30 MeV and 100 MeV, which differs from the
  standard LAT analysis. Filling the gap between the energy ranges
  where the Gamma-ray Burst Monitor (GBM) and LAT operate is important
  to better constrain the high-energy spectra of GRBs.
  The LAT Low-Energy (LLE) technique is described, first performance
  studies are presented, as well as preliminary spectral re-analyses of two
  {\it Fermi} GRBs.
\end{abstract}

\maketitle

\thispagestyle{fancy}

\section{INTRODUCTION}

Since the launch in June 2008, the Gamma-ray Burst Monitor (GBM)
onboard {\it Fermi} detected over 300 Gamma-Ray Bursts (GRB), 12 of
which were detected 
and studied above 100 MeV with the {\it Fermi} Large Area Telescope (LAT). 
The GBM enables GRB spectroscopy from 8 keV to 40 MeV, and the
standard LAT analysis is performed using events of energies
above 100 MeV only.

Here we propose a non-standard LAT analysis which
allows to recover the prompt emission from GRB between $\sim$30 MeV and 100
MeV, joining the two instruments' energy ranges. This so-called LAT
Low-Energy (LLE) technique also
improves the photon statistics above 100 MeV, and will allow us to
better define the GRB prompt emission spectral characteristics. It is
important to note that this analysis is based on non-public data and
software, and is still under improvement and calibration at the time of
this proceeding. 

Section~\ref{sect2} describes the LLE technique objectives and
principles. Peformances studies are presented in section~\ref{sect3}. Preliminary
re-analyses of the bright GRB 080916C and GRB 090510 are
presented in section~\ref{sect4}.

\section{THE ``LAT LOW-ENERGY'' TECHNIQUE}
\label{sect2}
\subsection{Motivation}

\begin{figure}
\begin{center}
  \includegraphics[width=0.9\linewidth,angle=0]{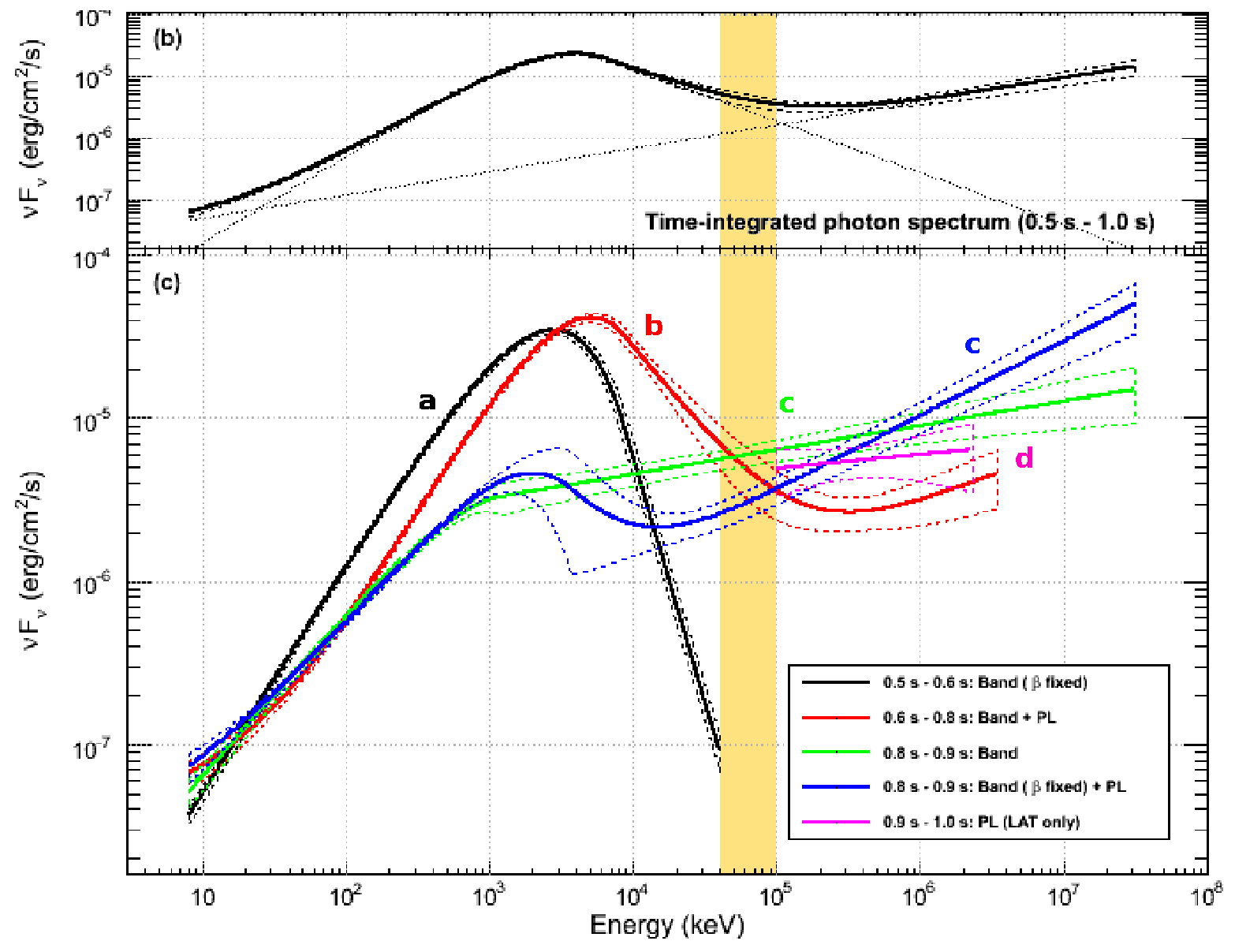}
\end{center}
\caption{GRB 090510 spectral energy distribution \cite{ApJ090510}. \\
  {\bf Top:} the time-integrated spectrum clearly shows an additional power-law
  component. {\bf Bottom:} time-resolved spectroscopy. In bin 'a' no
  signal was detected in the LAT. Bins 'b' and 'c' show an additional
  power-law component. No signal is detected in the GBM after 0.9 s
  (bin 'd'). The 40 -- 100 MeV gap is highlighted.}
\label{fig:090510_prompt_SED}
\end{figure}

The LAT standard analysis is based on a sophisticated reconstruction and
classification procedure \cite{LATpaper}. The probability for each
event to be a photon is estimated using the event topology and the
reliability of its reconstruction.

The characteristcs of the particle backgrounds below 100 MeV is still
under study and makes this classification more complicated. Moreover,
the LAT effective area suffers from systematic effects which are not
well determined yet in this energy range. Although the standard LAT
analysis currently starts at 100 MeV, public datasets and responses
should allow for analyses starting at energies lower than 100 MeV for
any kind of source in the future.

On the one hand, faint or soft GRB emissions (e.g. bin 'a' of GRB 090510, see
fig.~\ref{fig:090510_prompt_SED}) are not detected so far above 100
MeV, while they can exhibit additional components which become
dominant above few tens of MeV. Their signal above this energy can be
rejected by the usual quality 
selections or be intrinsically too faint. An improved spectral analysis
above 40 MeV would reveal any kind of feature, e.g. a
faint additional component, a cutoff, or a simple power-law behaviour. The
use of the present technique should improve the statistics above 100
MeV as well, although it is not yet clear whether it will
substabtially improve the sensitivity of the spectral analysis.

On the other hand, bright spectra (e.g. bin 'b' to 'd' of GRB 090510,
see fig.~\ref{fig:090510_prompt_SED}, or GRB
080916C~\cite{Sci080916c}) are currently reconstructed above 100
MeV. Adding a LAT low-energy signal to the available
standard dataset would help to better constrain the
spectral parameters of such features (see section~\ref{sect4}).

\subsection{Principle}

\begin{figure}
\begin{center}
\includegraphics[width=0.8\linewidth,angle=0]{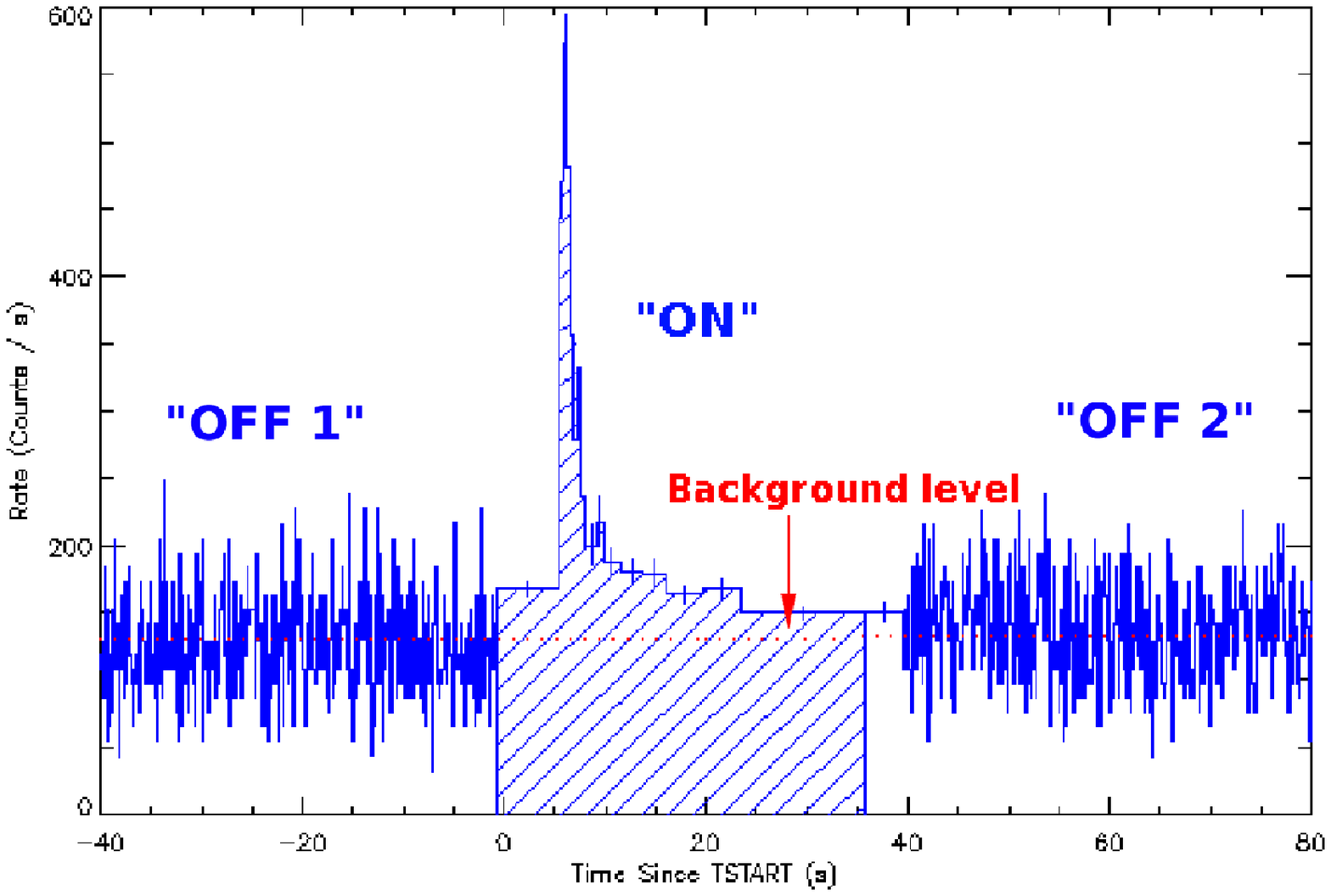}
\includegraphics[width=0.8\linewidth,angle=0]{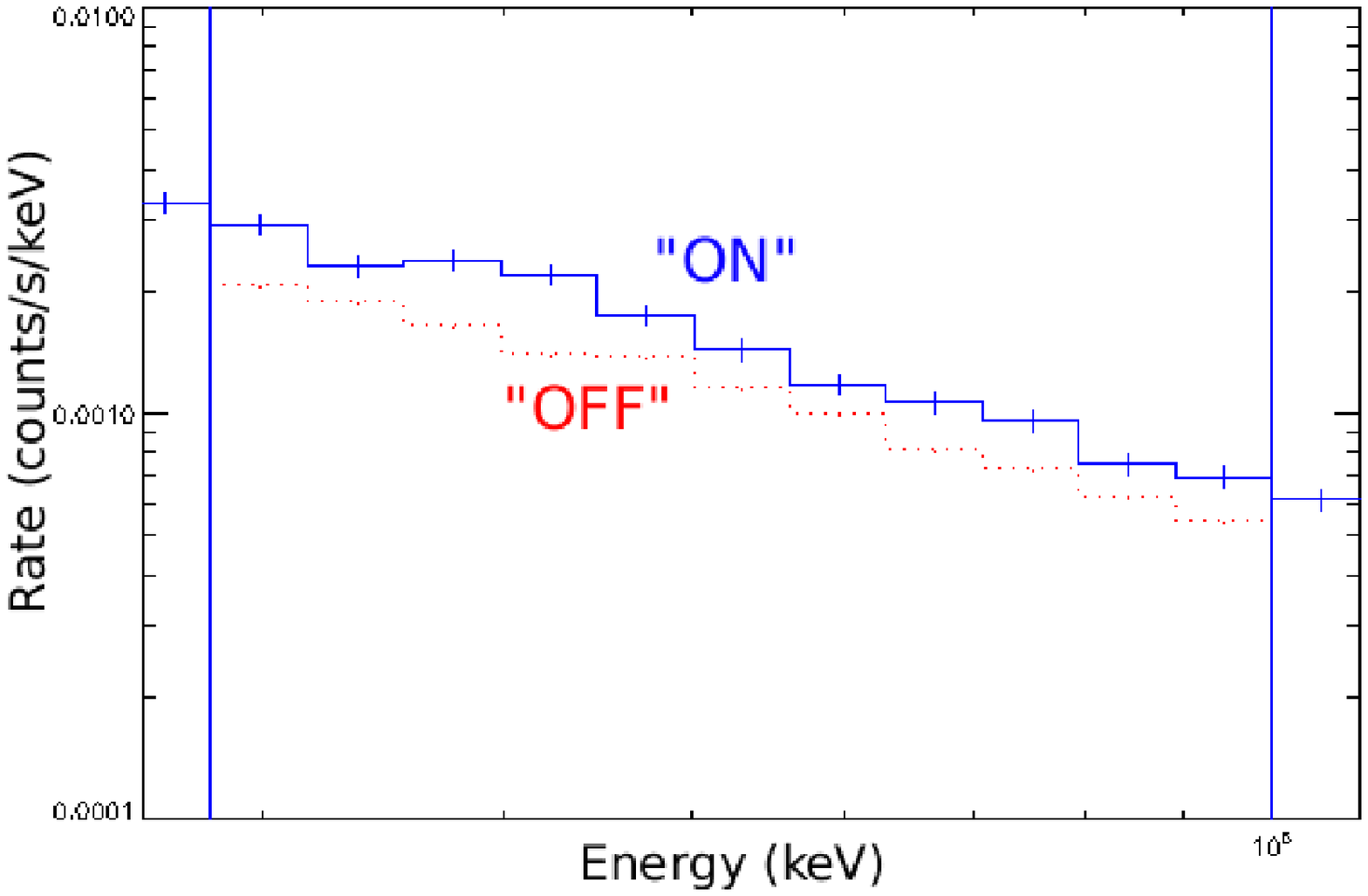}
\end{center}
\caption{GRB 080916C prompt emission between 30 MeV and 100 MeV. {\bf
    Top:} the lightcurve is used to define ``ON'' and ``OFF''
    emissions. The ``OFF'' rate is extrapolated to estimate the
    background level in the ``ON'' interval. {\bf Bottom:} signal and
    extrapolated background spectra in the ``ON'' interval.}
\label{fig:on_off}
\end{figure}

\begin{figure}
\begin{center}
\includegraphics[width=0.8\linewidth,angle=0]{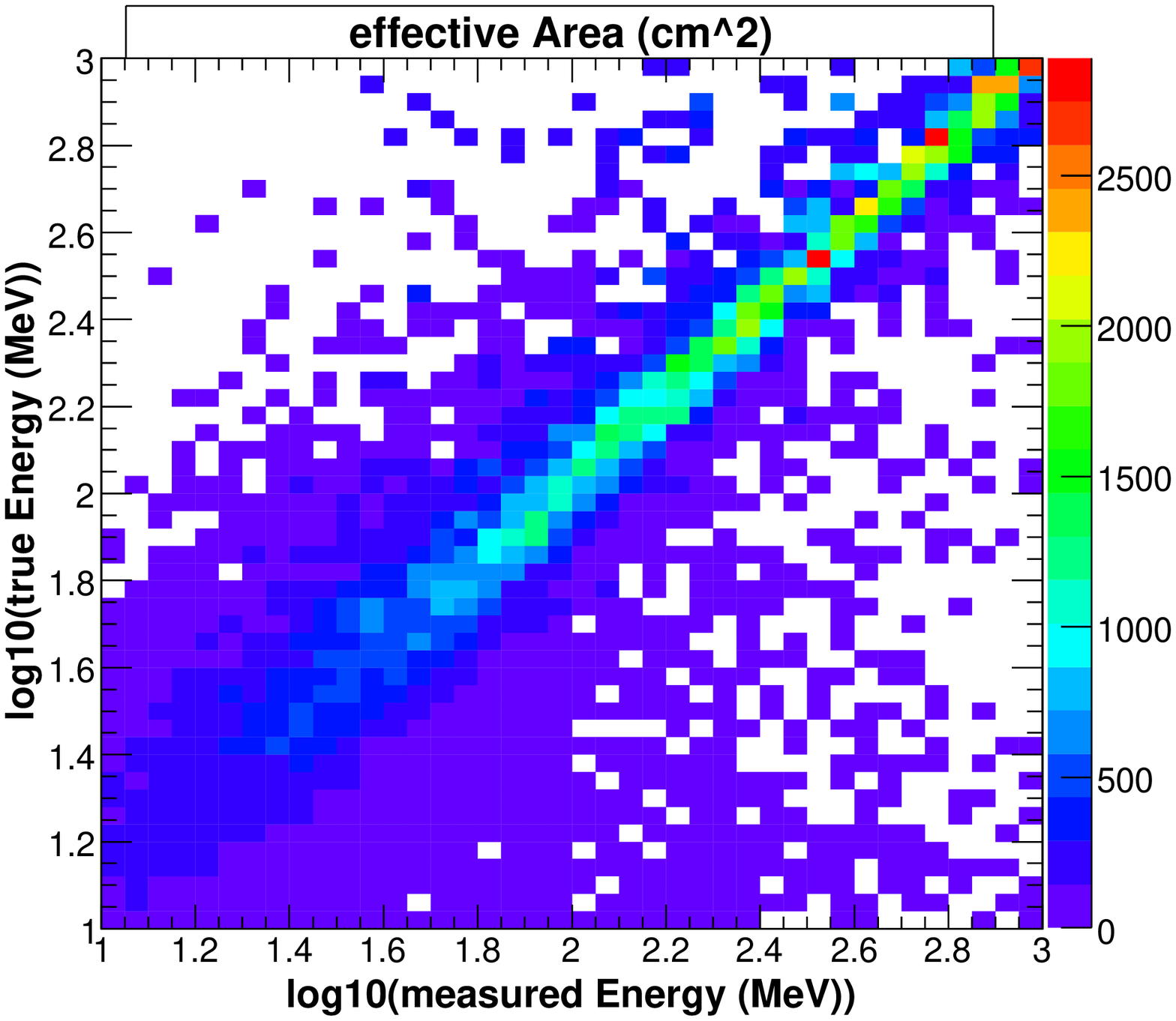}
\includegraphics[width=0.8\linewidth,angle=0]{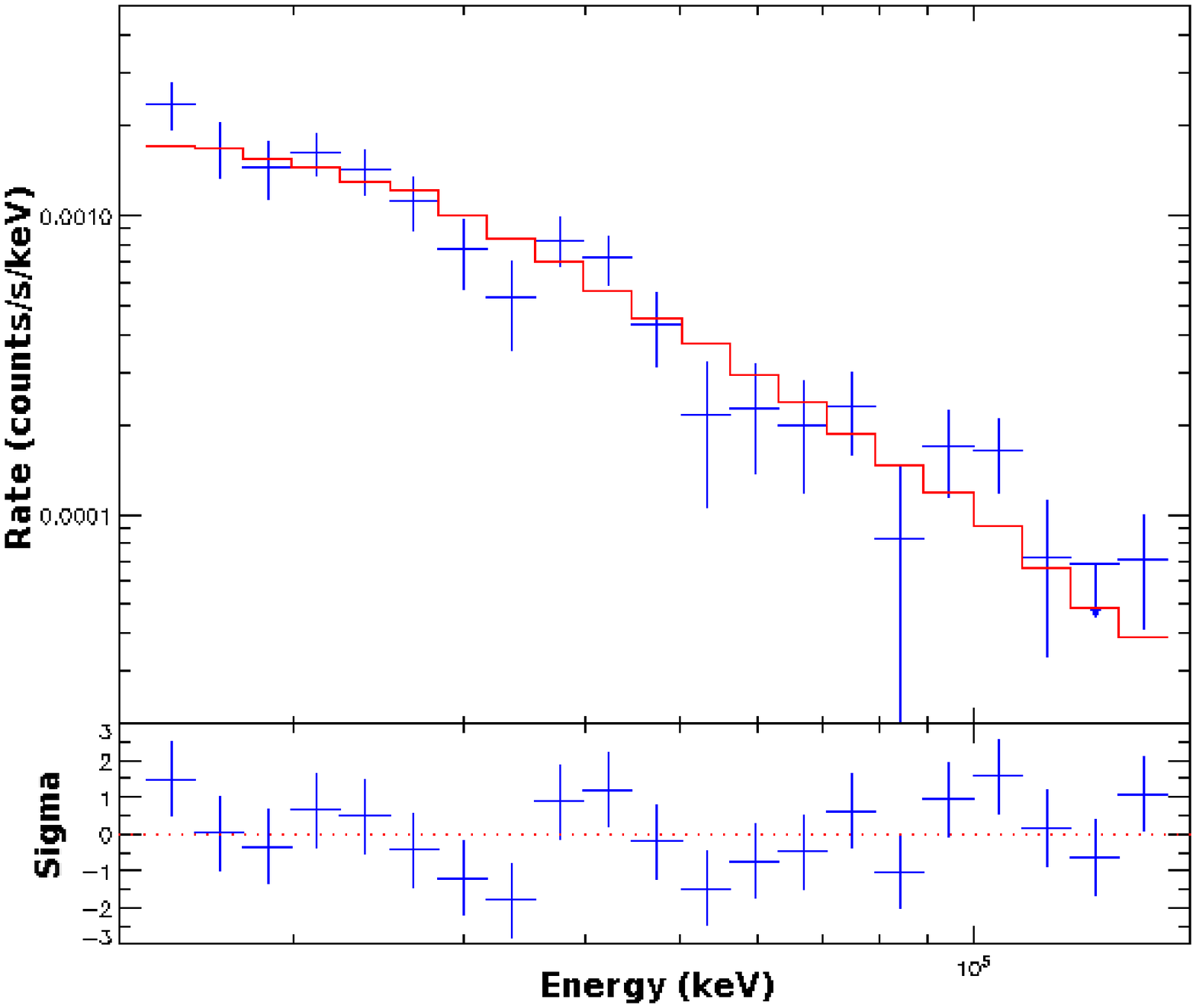}
\end{center}
\caption{{\bf Top:} DRM produced for GRB
  080916C LLE spectral analysis in the 10 MeV -- 1 GeV energy range.
  {\bf Bottom:} GRB 080916C counts spectrum between 30 MeV and 200 MeV (points)
  with folded spectrum (line) and residuals (bottom panel).}
\label{fig:drm_spec}
\end{figure}

The LLE technique makes use of the analysis software RMfit, which is
commonly used for all GBM spectral and temporal analyses. It consists
of the forward-folding analysis of a background-subtracted binned
event rate,
using a Detector Response Matrix (DRM) for model-folding
(fig.~\ref{fig:drm_spec}). A temporal selection defines the ``ON''
interval containing the emission and the ``OFF'' interval(s) used to
extrapolate the background rate in the ``ON'' interval, in each energy
bin (fig.~\ref{fig:on_off}).

In contrast with the standard analysis, the LLE event selection
does not include strict quality criteria. We consider every event
passing the onboard GAMMA filter~\cite{LATpaper} and for which at
least one track could be found in the tracker.  Many of these events do
not have a reliable direction, therefore no ``region of interest'' can
be defined around the source position in the sky. A binning is
applied on the measured energy (see section~\ref{sect32}). 

Once the ``ON'' and ``OFF'' intervals have been selected, the
background rate time profile is fitted with a polynomial function in
each energy bin, which yields an estimate of the background spectrum
in the ``ON'' interval. The background spectrum can vary with time so
the ``OFF'' intervals can not be chosen too far from the ``ON''
interval, nor too close nor short to allow for a good fit
precision. The same precautions are usually taken for spectral
analyses of GBM data.

The DRM is built from a dedicated extensive simulation of a bright
point source using the \textsc{gleam} software \cite{gleam}. 
The observation conditions reproduce those of the GRB of
interest. In particular, the inclination angle of the source and
livetime should be the same since both affect the effective area. The
LLE selections are applied to the dataset produced, the remaining
events are binned in true and measured energy. The resulting distribution
is converted to area units (fig.~\ref{fig:drm_spec}). The effect of
the simulated spectral shape and of the true energy binning on the
reconstruction quality are yet under investigation.

\section{PERFORMANCES STUDIES}
\label{sect3}
\subsection{Cuts efficiencies}

\begin{figure}
\begin{center}
\includegraphics[width=0.9\linewidth,angle=0]{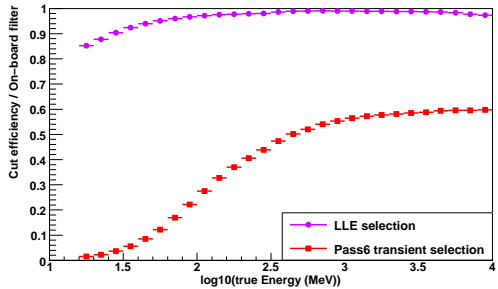}
\end{center}
\caption{Fraction of events from a large set of simulated photons passing the
  standard Pass6 ``transient class'' quality selections (squares) and LLE relaxed
  selections (circles), both normalized to the onboard GAMMA
  filter.}
\label{fig:efficiency}
\end{figure}

The LLE selection consists of the onboard GAMMA filter and the
identification of at least one track in the tracker. This latter
criterion ensures a better correlation between the true and measured
energy (see section~\ref{sect32}).

These relaxed selections greatly improve the photon statistics at low
energies with respect to the standard Pass6 ``transient class'' quality selections, and improve
the statistics at high energies as well, as shown in
fig.~\ref{fig:efficiency} : the effective area is increased by a
factor of $\sim$50 at 30 MeV, 4.5 at 100 MeV, and still by more than a
factor 2 above 1 GeV.

\subsection{Energy resolution}
\label{sect32}

\begin{table}
\small
\begin{center}
\caption{
  LLE energy measurement resolution for simulated photons of
  inclination angle $\theta\,<\,$40$^\circ$ (top) or
  40$^\circ\,<\,\theta\,<\,$70$^\circ$ (bottom). Energies are in
  MeV. True energy $E_{MC}$ lines of width 4 MeV were selected. The
  resolution is here defined as RMS/$<$E$_{mes}>$ and the error as
  ($<$E$_{mes}>$ - E$_{MC}$)/E$_{MC}$.}
\label{tab:energy_res}
\begin{tabular}{|c|c|c|c|c|}
\hline
E$_{MC}$ & $<$E$_{mes}>$ & RMS & resolution & error \\
\hline
30 & 27 & 10 & 37\% & -10\% \\
50 & 45 & 16 & 36\% & -10\% \\
100 & 90 & 27 & 30\% & -10\% \\
500 & 490 & 70 & 14\% & -2\% \\
\hline
30 & 30 & 14 & 47\% & 0\% \\
50 & 44 & 18 & 41\% & -12\% \\ 
100 & 85 & 34 & 40\% & -15\% \\
500 & 470 & 80 & 17\% & -6\% \\
\hline
\end{tabular}
\end{center}
\normalsize
\end{table}

The energy estimate used for this analysis is different from the
standard one. It is based on energy
depositions in both the tracker and the
calorimeter. For low-energy events (below 100 MeV) which barely reach the
calorimeter, the tracker mainly contributes to the energy
estimate. Hence the requirement of at least one reconstructed track.
The choice of this energy estimate may not be final, though it nicely
correlates to the true energy for simulated datasets (see
fig.~\ref{fig:drm_spec}).

A study of the energy resolution from this estimate has been
performed, using large samples of simulated photons (see
table~\ref{tab:energy_res}). No very large bias is found, and the
energy resolution is reasonable even at low energies (50\% above
$<$30 MeV, $<$40\% above 50 MeV).

\subsection{Reconstruction capabilities}

\begin{figure}
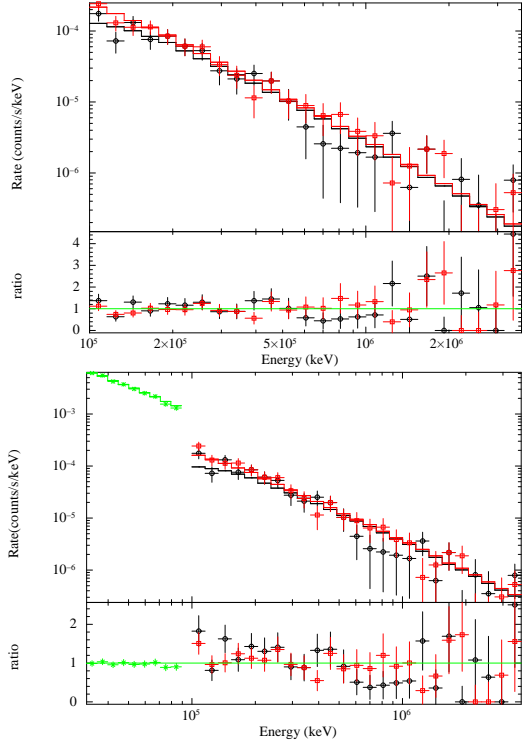

\begin{center}
\includegraphics[width=0.6\linewidth,angle=270]{fig/normPL_cstat_front_back_22.eps}
\includegraphics[width=0.6\linewidth,angle=270]{fig/normPL_cstat_LLE_front_back_22.eps}
\caption{Counts spectra of a simulated bright GRB with folded model and
ratio of the signal to the folded model. {\bf Top:} only standard Pass6
transient events above 100 MeV are used, events converting at the
front and back of the tracker are separated\cite{ApJ080825C}. {\bf
  Bottom:} LLE events between 30 MeV and 100 MeV are added to the
previous dataset. The ratio panel shows a good agreement between both
sets, and the fit is dominated by the LLE data.} 
\label{fig:sim_080916c}
\end{center}
\end{figure}

\begin{table}[b]
\begin{center}
\caption{Spectral parameters of the simulated GRB spectrum, and
  reconstructed spectra using standard Pass6 transient events
  only, or LLE events as well. Only statistical errors are
  shown. Though the use of LLE data reduces the uncertainties, a
  systematic error appears.}
\label{tab:sim_080916c}
\begin{tabular}{|c|c|c|}
\hline
Spectrum & $N_0$ & $\beta$ \\
 & (10$^{-9}$ ph.cm$^{-2}$.s$^{-1}$) & \\
\hline
input & 1.19 & 2.1 \\
\hline
fit without LLE & 1.36 $\pm$ 0.13 & 2.21 $\pm$ 0.06 \\
fit with LLE & 1.60 $\pm$ 0.12 & 2.02 $\pm$ 0.02 \\
\hline
\end{tabular}
\end{center}
\end{table}

To study the reconstruction capabilities of the LLE technique, a bright
GRB was simulated as a point source with a simple power-law spectrum :
$N(E) = N_0 (E/E_0)^{-\beta}$. No background was added to the data and
this spectrum was analyzed using the standard tool
XSPEC (HEASARC). The reconstruction was performed using only the
standard Pass6 transient data (above 100 MeV), or using both the LLE
(30 MeV to 100 MeV) and Pass6 transient data. Both reconstructions
yield similar results (table~\ref{tab:sim_080916c}), with a smaller 
uncertainty on the index $\beta$ when using LLE data. A systematic
error yet appears, which will be investigated after the technique has
been fully calibrated.

\section{PRELIMINARY RE-ANALYSES}
\label{sect4}

New analyses of two bright bursts were performed and yield encouraging
results.

\begin{figure}
\begin{center}
\includegraphics[width=0.8\linewidth,angle=0]{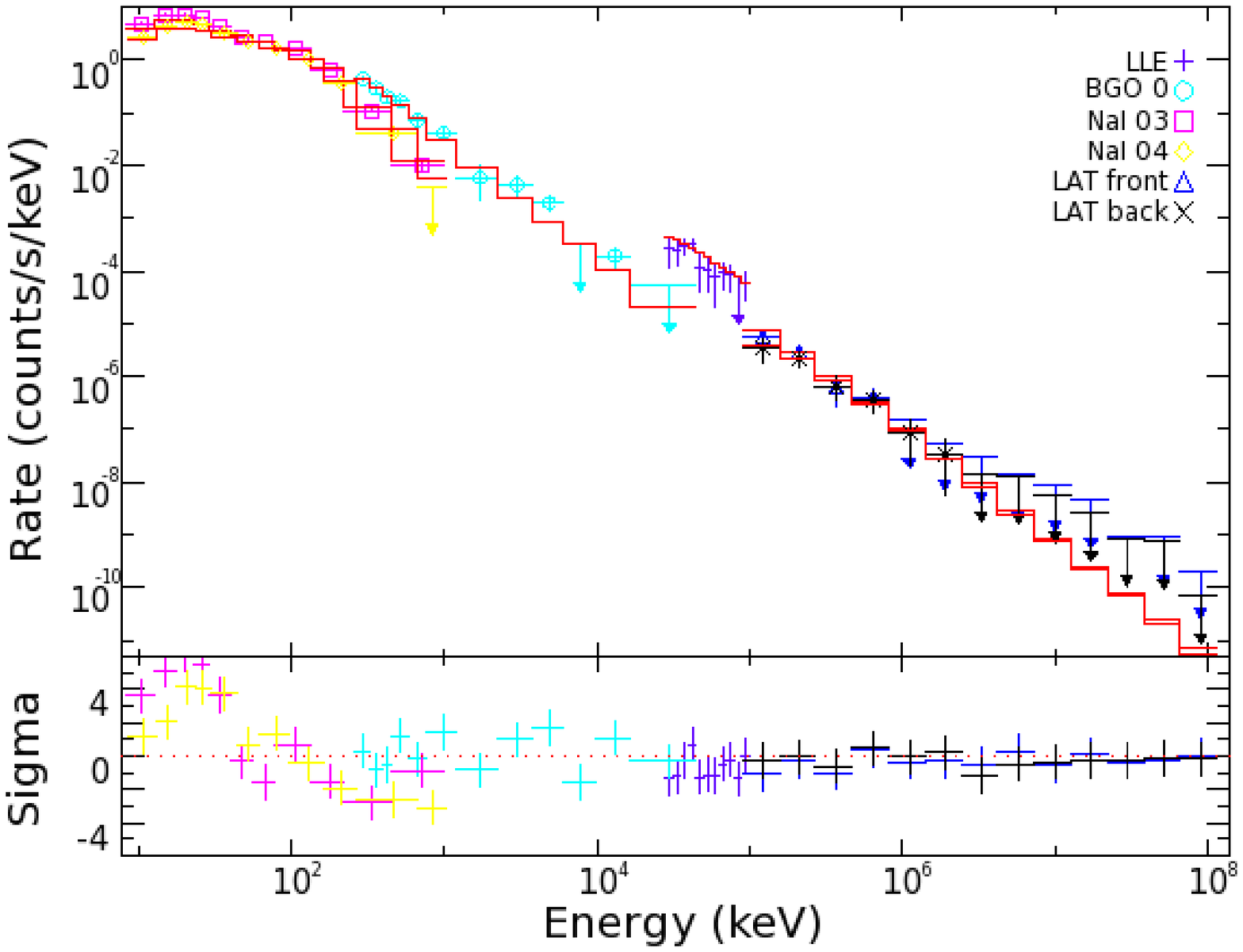}
\includegraphics[width=0.8\linewidth,angle=0]{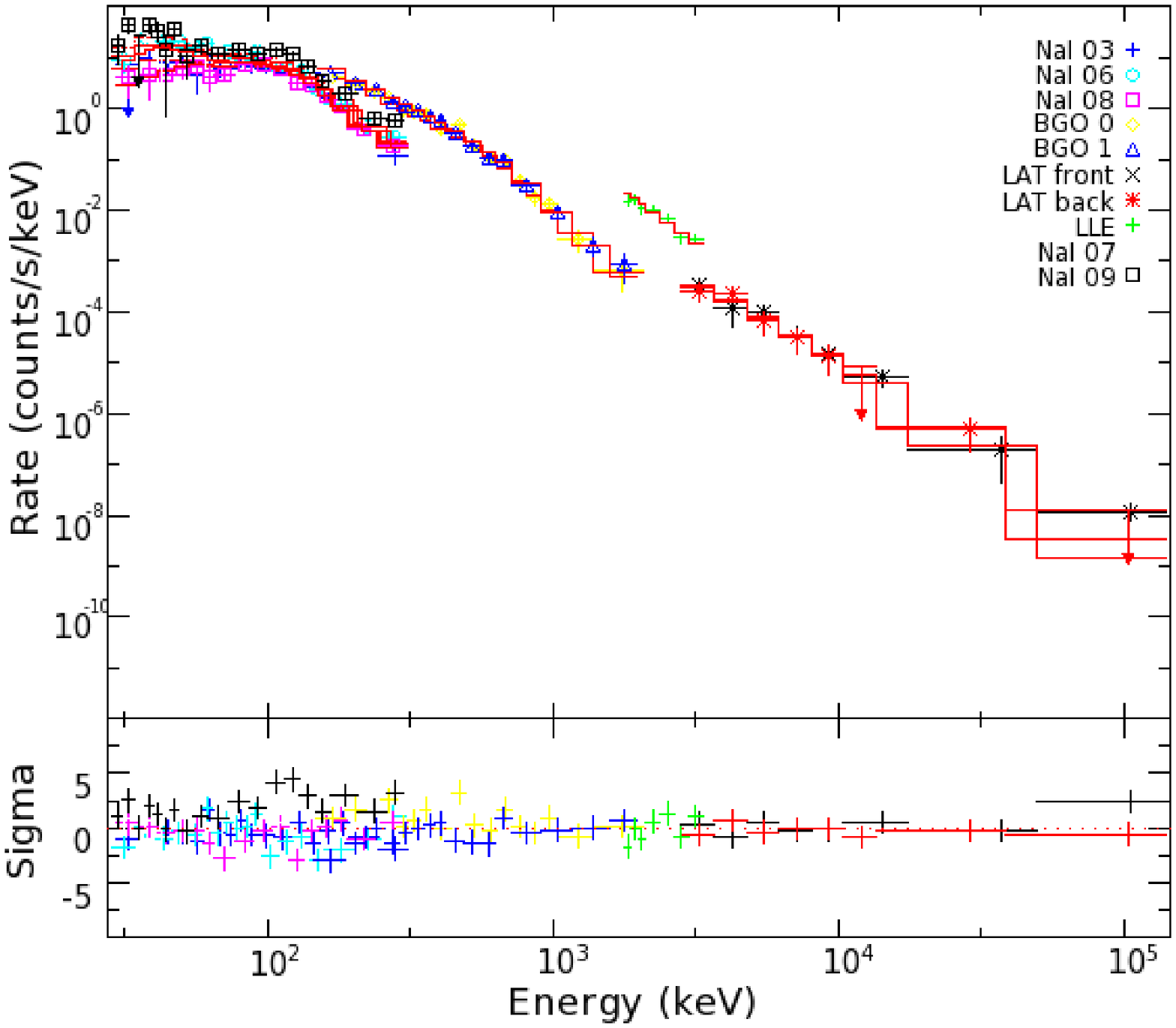}
\end{center}
\caption{{\bf Top: }GRB 080916C counts spectrum with folded model and
  residuals. GBM and LAT standard transient data were fitted. LLE
  dataset was superimposed to the spectrum.
  {\bf Bottom: }GRB 090510 time-integrated counts spectrum with folded model and
  residuals. GBM data, LAT Pass6 transient data (above 100 MeV) and
  LLE data (30 MeV -- 100 MeV) were fitted together by a Band
  function with an additional power-law component.}
\label{fig:ana_080916c}
\end{figure}

GRB 080916C time-integrated spectrum was fitted using only GBM and
standard transient events (above 100 MeV)
(fig.~\ref{fig:ana_080916c}). The LLE dataset was added but not used
for the fit. The good residuals in the LLE energy range (30 MeV -- 100
MeV) show the good agreement between this technique and the standard procedure.

GRB 090510 time-integrated spectrum was fitted using all datasets :
GBM, LAT Pass6 transient, LLE. A Band function with an additional
power-law component yields the best fit, like in the standard analysis
\cite{ApJ090510}. The high-energy component is more significant
when using LLE data ($N_\sigma$ = 8.9 instead of 5.6 without
LLE). The spectral evolution observed in this bright burst's prompt
emission will certainly better show up if the LLE data are used in the
time-resolved spectroscopy. 

\newpage
\section{CONCLUSION}

The LLE technique presented here can be used in principle for any kind of flaring source
: GRB prompt emission, AXP, pulsars (see \cite{MBurgess} for
details), etc.

In particular it appears to be very promising for GRB
prompt emission spectral analyses, revealing or better defining GRB
spectral features above 30 MeV. 

The validation study of these analyses is yet
still ongoing (acceptance and energy calibration, systematics
effects). The reconstruction of a simulated spectrum has shown
a bias, such effects have to be characterized and understood. The
performances that are yet measured from sets of simulated photons have
to be confronted to data, e.g. using Vela on-pulse emission as a pure sample
of real photons.

\end{document}